\documentclass[reprint,superscriptaddress,amsmath,amssymb,aps,prc,showpacs]{revtex4-1}

\usepackage{array}
\usepackage{tabularx}
\usepackage{graphicx}
\usepackage{dcolumn}
\usepackage{bm}
\usepackage{natbib}

\newcommand{\etal}{et \emph{al. }}

\begin{document}

\title{Description of He isotopes using the\\ particle-particle random-phase approximation model}
\author{B.~Laurent}
\email{Corresponding author: benoit.laurent@cea.fr}
\affiliation{%
 CEA, DAM, DIF F-91297 Arpajon Cedex, France
}
\author{N.~Vinh Mau}
\affiliation{%
 CEA, DAM, DIF F-91297 Arpajon Cedex, France
}
\affiliation{%
 Institut de Physique Nucl\'{e}aire, IN2P3-CNRS, Universit\'{e} Paris-Sud, F-91406 Orsay Cedex, France
}

\date{\today}

\begin{abstract}
The two-neutron RPA model has been used to describe helium isotopes with N (the neutron number) = 4, 6, 7, 8 in their ground and excited states. The properties of all these isotopes are given by a single system of equations so that their properties are interdependent. Since the ground states of $^{9}$He and $^{10}$He are still not well established, we have looked how the properties of $^{9}$He induce the properties of the other isotopes, energies and wave functions. Our results suggest an inversion of $2s$-$1p_{1/2}$ shells in $^{9}$He. The corresponding ground states of $^{9}$He and $^{10}$He are slightly unbound and respectively $1/2^+$ and $0^+$ states while the $1/2^-$ and $0^+$ seen in experiments appear to be excited states. With this assumption on $^{9}$He, we get not only a nice picture of $^{10}$He but also a very good two-neutron separation energy in $^{8}$He.
\end{abstract}


\maketitle

\section{Introduction}\label{sec:intro}
Despite their small number of nucleons, the light nuclei of the nuclide chart provide a unique opportunity to investigate complex phenomena such as halo nuclei, cluster configurations, shell inversions, and give access to nuclear systems beyond the neutron dripline~\cite{jon04}. This region thus challenges the theory in its capacity to reproduce the wide variety of these phenomena. In that sense, the unbound helium isotopes, $^{9}$He and $^{10}$He have been
extensively studied both experimentally~\cite{set87,boh88,oer95,chen01,rog03,for07,gol07,alf07,joh10,mat12} 
and theoretically~\cite{pop85,war92,pop93,kit93,chen01,ots01}. 
The first experiments on $^{9}$He showed a $1/2^-$ state in the range of 1-2~MeV interpreted
as its ground state~\cite{set87,oer95,boh88,boh99}. This property of a $1/2^-$ ground
state was also the result of most of the calculations~\cite{ogl95,aoy97}. A latter double proton knockout experiment however shows the existence of a lower state~\cite{chen01}, close to the $n+^{8}$He threshold, compatible with a possible $1/2^+$ state. This assignment has been comforted by a shell model calculation~\cite{chen01}. It has been identified in a further experiment~\cite{for07} but is still disputed in recent experiments~\cite{joh10,alf07,mat12}.
There is also an ambiguity concerning the $0^+$ resonance seen experimentally in $^{10}$He and interpreted usually as the ground state while a suggestion that it is an excited state has been proposed by S. Aoyama~\cite{aoy02,aoy03} but has not been confirmed by other calculations.

Such a situation of a $1/2^+$ ground state with a $1/2^-$ excited
state reveals an inversion of the $2s$ and $1p_{1/2}$ shells when $^{9}$He is described as a neutron added to a core
of $^{8}$He. Such an inversion is well known in $^{11}$Be~\cite{deu68,nun96}, $^{10}$Li~\cite{vin96,nun96}
and $^{13}$Be~\cite{nak10}. In $^{10}$Li and $^{13}$Be this inversion was first
suggested theoretically in order to reproduce the two-neutron
separation energy in $^{11}$Li and $^{14}$Be respectively~\cite{vin96,lab99,bla10}, and then experimentally confirmed~\cite{for07,nak10,sid12}. These two-neutron
separation energies were calculated in a two-neutron RPA model where
$^{11}$Li and $^{14}$Be were considered as two neutrons
added to an inert core of $^{9}$Li and $^{12}$Be respectively. In the present
paper, the same two-neutron RPA model depicting $^{10}$He as
two neutrons added to an inert core of $^{8}$He is used. This model
gives one system of equations, describing both $^{10}$He and
$^{6}$He; the latter being described as two-neutrons subtracted from a $^{8}$He core.
Finally these calculations provide, among other properties, the two-neutron separation
energies ($S_{2n}$) of $^{8}$He and $^{10}$He, the energies and wave functions of the $0^+$ states in $^{6}$He and $^{10}$He as well as informations on the composition of the $^{8}$He core wave function.

In section \ref{sec:model} the two-neutron RPA
model and the inputs used in the calculation are briefly presented. The results are given in section
\ref{sec:results} and the conclusions are compiled in the final section.

\section{The model}\label{sec:model}
Only a brief summary of the used two-neutron RPA model is given. More details can be found in reference~\cite{pach02}. 
A core of $^{8}$He in its $0^+$ ground state is assumed which, in the Hartree-Fock model, is a
closed shell nucleus with the last neutrons filling the $1p_{3/2}$
shell. Two neutrons are then added to, or subtracted from this core to describe
$^{10}$He and $^{6}$He as eigenstates of a single system of
equations. Eigenvalues and eigenvectors of the RPA matrix
yield the energies of $^{6}$He and $^{10}$He referred to the core
energy and the two-body amplitudes in $^{6}$He and $^{10}$He with
the following definition for a solution of rank n:
\begin{eqnarray}
    X^{(n)}(a) & = & \langle \psi_n(^{10}He) |  A^+_a  | \psi_0(^{8}He)\rangle \label{eqs:wfampX},\\
    Y^{(n)}(a) & = & \langle \psi_n(^{6}He) |  A_a  | \psi_0(^{8}He)\rangle  \label{eqs:wfampY},
\end{eqnarray}
with the orthonormalization relations:
\begin{eqnarray}
    \sum_{\substack{unoccupied\\states}} |X(a)|^2  & - & \sum_{\substack{~~~occupied\\~~~states}} |X_(a)|^2 = 1 \nonumber,\\
    \sum_{\substack{~~~occupied\\~~~states}} |Y(a)|^2 & - & \sum_{\substack{unoccupied\\states}} |Y(a)|^2   = 1 \label{eqs:wfamportho},
\end{eqnarray}
where $A^+_a$ and $A_a$ are respectively the creation and annihilation operators of two neutrons in state
$\bm{a}$ formed of two occupied or unoccupied states in the
Hartree-Fock core. $\psi_0(^{8}He)$ is the correlated wave function of the
$^{8}$He core. Then, for example, a non zero amplitude $\bm{X}$ for $\bm{a}$,
representing two occupied states, means that in the core wave
function there is a configuration with two holes in that state. Similarly, if $\bm{Y}$ is non zero for a state $\bm{a}$ formed of two
unoccupied states, then the $^{8}$He wave function contains a configuration with two particles in that state $\bm{a}$. The energies of the $^{6}$He and $^{10}$He ground states give directly the two-neutron separation energies in $^{8}$He and $^{10}$He respectively.

To solve the RPA equations, we need the single neutron energies in the field of the core and the neutron-neutron interaction which we take as the Gogny DIS effective interaction~\cite{dech80,berg91}. For the neutron states in the field of $^{8}$He, we take a Saxon-Woods potential with a surface term which simulates the contribution to the one body potential of neutron-phonons couplings~\cite{vin95,vin96}. It is written as:
\begin{eqnarray}
    V_{nc}(r) = -V_{NZ}\left(f(r)-0.44r_0^2(\textbf{l.s})\frac{1}{r}\frac{df(r)}{dr} \right)\nonumber\\
             +16a^2\alpha_n\left(\frac{df(r)}{dr}\right)^2,\\
    with~f(r) = \left( 1 + exp\left[\frac{r-R_0}{a}\right] \right)^{-1}. &
\end{eqnarray}
The following parameters suggested by Bohr and Mottelson~\cite{bohr}, are used:
\begin{eqnarray}
  & V_{NZ} = \left(51 - 33 \frac{N-Z}{A_c}\right)~MeV, \\
  & R_0  =  r_0A_c^{1/3},~r_0 = 1.27~fm,
\end{eqnarray}
where N, Z, $A_c$ refer to the core nucleus.
The diffusivity is set to a larger value, $a=0.75$~fm, to account for the diffuse
surface of such a light nucleus. It has been shown that neutron-phonon couplings are responsible of the inversion of$1p_{1/2}$ and $2s$ shells in $^{11}$Be~\cite{vin95,gor04}. Most of the contribution of such couplings comes from the low energy $2^+$ excited state in the core of $^{10}$Be. This inversion is also present in $^{10}$Li and $^{13}$Be where the cores of $^{9}$Li and $^{12}$Be have also low energy $2^+$ states with large B(E2). Such couplings are also responsible for an inversion of $2s$ and $1d_{5/2}$ shells in the light carbon isotopes and also due to $2^+$ phonons~\cite{lau12}. Since $^{8}$He has a $2^+$ state at 3.54 MeV~\cite{gri09}, we expect that neutron-$2^+$ phonon coupling will induce large modification of the neutron energies close to the Fermi surface and could be responsible for an inversion of $1/2^-$ and $1/2^+$ states in $^{9}$He. In our calculation we take the strength $\alpha_n$ of the surface term as a parameter fitted on known or assumed energy for states close to the Fermi 
surface as $1p_{3/2}$, $1p_{1/2}$ and $2s$ states. For higher states, the coupling effects are small and we take $\alpha_n = 0$. For $1p_{3/2}$, $\alpha_n$ is fitted to give $\epsilon(1p_{3/2})=-S_n(^{8}He)=-2.57$~MeV.
To solve the one-body equation we use a discretisation of unbound states with a box of 16~fm. This discretisation of continuum states has been tested in $^{6}$He~\cite{vin05} and shown to bring an under-estimation of pairing energy of few percents only. Working with discrete states means that we get the position of the resonances but do not get any width.

\section{Results}\label{sec:results}
All experiments agree on the presence in the $^{9}$He spectrum of a
$1/2^-$ state in the vicinity of 1-1.27~MeV energy range. This state was, up to
L. Chen and collaborators' experiment~\cite{chen01}, considered as the ground state of
$^{9}$He. This was attested by several shell model calculations.
However L. Chen \etal experiment and most of the recent ones agree that the
$1/2^-$ state is an excited state and that the ground state is an
unbound virtual $1/2^+$ state, with an energy very close to
the $n+^{8}$He threshold. On the other end, there is an ambiguity in
the interpretation of experiments on $^{10}$He. Indeed until
recently the resonance found at about 1~MeV above the $2n+^{8}$He
threshold was considered to be the ground state of $^{10}$He. However
there is a recent suggestion by~\cite{aoy97} that this resonance
corresponds to a $0^+$ excited state and that the ground state is
close to the threshold. The use of the two-neutron RPA model assuming a
core of $^{8}$He gives the opportunity to make a simultaneous study
of $^{6}$He, $^{8}$He, $^{9}$He and $^{10}$He where the choice of
parameters defining $^{9}$He states determines the states of the
other three nuclei without any further fitting possibility. The calculations
proceed as follows: the $1p_{3/2}$ energy is set to the experimental
energy of -2.57~MeV. We put the $1p_{1/2}$ state close to the measured energy of the $1{1/2}^-$ state seen in all experiments and vary the energy of the $2s$ state. The resolution of the RPA equations gives us the corresponding properties of the other nucleus. The results are presented in table \ref{tab:1.25} for
$\epsilon(1p_{1/2})=1.25$~MeV, which is the value given in most of
the experiments' analysis and in table \ref{tab:0.92} for $\epsilon(1p_{1/2})=0.92$~MeV, which
is not excluded by experiments. In these tables are given the results for $S_{2n}(^{8}He)$, $S_{2n}(^{10}He)$ and $E(0^+_2)$. The latter corresponds to the
energy of the first excited $0^+$ state in $^{10}$He relative to
the $n+n+^{8}$He threshold. For
$\epsilon(1p_{1/2})=1.25$~MeV and $\epsilon(2s)\approx0.3-0.4$~MeV,
$^{10}$He is unbound with a small value of $S_{2n}$ in agreement
with the assumption that the resonance at about 1~MeV is not the
ground state but a $0^+$ excited state. Yet the energy of this
excited state is too high. Furthermore $S_{2n}(^{8}He)$ is slightly
too large compared to the experimental measured value of 2.14~MeV~\cite{audi03}. If we assume the $2s$ shell to be higher than the 
$1p_{1/2}$ one, for any value of $\epsilon(p_{1/2})$, the results are in complete disagreement with measurements.

\begin{table}
  \caption{\label{tab:1.25} Results in terms of $\epsilon(2s)$ when $\epsilon(1p_{1/2})=1.25$~MeV}
   \centering
  \begin{ruledtabular}
  \begin{tabular}{ldddd}
  $\epsilon(2s)$    &  0.78  &  0.45  &  0.29  & 0.192 \\ \hline
  $S_{2n}(^{8}He)$  &  2.52  &  2.48  &  2.48  & 2.43  \\
  $S_{2n}(^{10}He)$ & -1.33  & -0.53  & -0.11  & 0.12  \\
  $E(0^+_2)$        &  2.3   &  2.3   &  2.3   & 2.3   \\
  \end{tabular}
  \end{ruledtabular}
\end{table}

\begin{table}
  \caption{\label{tab:0.92} Results in terms of $\epsilon(2s)$ when $\epsilon(1p_{1/2})=0.92$~MeV}
   \centering
  \begin{ruledtabular}
  \begin{tabular}{ldddd}
  $\epsilon(2s)$    &  0.785  &  0.45  &  0.29  & 0.19 \\ \hline
  $S_{2n}(^{8}He)$  &  2.36   &  2.31  &  2.29  & 2.24  \\
  $S_{2n}(^{10}He)$ & -1.11   & -0.44  & -0.07  & 0.22  \\
  $E(0^+_2)$        &  1.6    &  1.4   &  1.4   & 1.4   \\
  \end{tabular}
  \end{ruledtabular}
\end{table}
On the other hand, in table \ref{tab:0.92} for $\epsilon(1p_{1/2})=0.92$~
MeV a general agreement is observed for all calculated quantities though the
most accurate are obtained for $\epsilon(2s) \approx 0.3$~MeV. This value of 0.92~MeV
 is slightly smaller than the values derived from experiments. However if one looks, for example, at the figure~1 in ref.~\cite{alf11}, a $p_{1/2}$ resonance at
about 0.9~MeV would give the same or even better agreement than 1.27~MeV. With these
energies of $2s$ and $1p_{1/2}$ states, the two-neutron separation
energy in $^{8}$He is very close to the measured value of 2.14~MeV
while the energy of the excited $0^+$ state in $^{10}$He is close to
the energy of 1.2~MeV deduced from experiments. The advantage of this
model is that, once the single neutron energies are fixed, the properties of all nuclei are given, without any possibility to
introduce further parameters to modify the results for one or the
other of the nuclei.

The RPA equations also yield the wave functions of $^{6}$He and
$^{10}$He through the two-neutron amplitudes defined in
Eqs.~(\ref{eqs:wfampX}-\ref{eqs:wfamportho}). In the case where
$\epsilon(1p_{1/2})=0.92$~MeV, $\epsilon(2s)=0.29$~MeV, the following RPA amplitudes of Eqs.~(\ref{eqs:wfampX}-\ref{eqs:wfamportho}) for the
$^{10}$He ground state are obtained as:
\begin{eqnarray*}
  X(2s^2)& = & 0.98,\\
  X(p_{1/2}^2)& = & -0.28,\\
  X(p_{3/2}^2) & = & -0.28.
\end{eqnarray*}
The main contribution comes from the $(2s)^2$ configuration for the
two last neutrons but the amplitude for the two neutrons added to the $(1p_{3/2})$ shell is quite large and means that in the core of $^{8}$He there is a non
negligible contribution of a 2 holes-2 particles configuration 
with the holes in the $p_{3/2}$-shell. 

For the $^{6}$He ground state, the obtained amplitudes are:
\begin{eqnarray*}
  Y(p_{3/2}^{-2}) & = & 1.08,\\
  Y(p_{1/2}^{-2}) & = & 0.21,\\
  Y(2s^{-2}) & = & -0.20.
\end{eqnarray*}
These values of the Y amplitudes indicate again a contribution of a 2
holes-2 particles configuration in the $^{8}$He core with the two particles in the
$1p_{1/2}$ or $2s$ shells. Then from these two results
it can be understood that $^{8}$He in its ground state is not a pure closed shell nucleus but has a contribution of
$(1p_{3/2})^{-2}-(1p_{1/2})^2$ or $(1p_{3/2})^{-2}-(2s)^2$ configurations. The $0^+$ excited
state of $^{10}$He appears to have mostly a $(1p_{1/2})^2$
configuration with some mixture of $(2s)^2$, with the respective RPA 
amplitudes of 0.96 and 0.25.

\section{Conclusions}\label{sec:discussion}
The difference between our work and previous theoretical works relies on two facts. First, we use a two-neutron RPA model which gives simultaneously properties of the $^{6}$He, $^{8}$He, $^{9}$He and $^{10}$He isotopes while usually calculations concern one of these nuclei only. Second, the only parameters of the model are the neutron energies for $2s$ and $1p_{1/2}$ states in $^{9}$He considered as a $^{8}$He + n system. The energy of the $1p_{1/2}$ neutron state is varied in the vicinity of the measured energy of the $1/2^-$ state seen in experiments. For the $2s$ state we vary its energy starting from very low positive energy. Then for each couple we look at the results of the RPA system of equations which gives interrelated informations on the other nuclei. We then look for the best overall results, namely which is the best case for which all known quantities are well reproduced, then see what are the implications for other properties. The results can be summarized as follows:
\begin{itemize}
\item in $^{9}$He the ground state should be an unbound $1/2^+$ state very close to the $n+^{8}$He threshold with an energy of 0.2-0.3~MeV, while the $1/2^-$ found experimentally close to 1-1.2MeV is an excited state. Therefore, we see in $^{9}$He the same inversion of $2s$-$1p_{1/2}$ shells as known in $^{11}$Be, $^{10}$Li and $^{13}$Be. We guess from our previous works that this inversion is due to the coupling of the neutron with the $2^+$ phonon known at 3.59~MeV in $^{8}$He.
\item the $^{10}$He has an unbound ground state very close to the $n+n+^{8}$He threshold and the resonance seen in experiments at 1.2~MeV and identified as the ground state in many theoretical or experimental works, is the first $0^+$ excited state.
\end{itemize}

Therefore our results give support to the suggestions of L. Chen for $^{9}$He and S. Aoyama for $^{10}$He which are still considered as doubtful in recent theoretical as well as experimental studies.

Furthermore the RPA amplitudes obtained for $^{6}$He and $^{10}$He show that the
core of $^{8}$He is not a pure closed $(1p_{3/2})$ shell nucleus but has a qualitatively important mixture of
$(1p_{3/2})^{-2} (1p_{1/2})^2$ and $(1p_{3/2})^{-2} (2s)^2$ configurations. The
wave functions of $^{6}$He and $^{10}$He show also some mixtures of various
configurations with still a strong component on two neutron holes in the $p_{3/2}$-shell
for $^{6}$He and on two neutrons in the $2s$-shell and $p_{1/2}$-shell for respectively the
ground state and the $0_2^+$ excited state in $^{10}$He.

\bibliography{bib_he_cor}

\end{document}